\documentclass[%
reprint,
 amsmath,amssymb,
 aps,
prb,
showpacs,
]{revtex4-1}

\usepackage{graphicx}% Include figure files
\usepackage{dcolumn}% Align table columns on decimal point
\usepackage{bm}% bold math
\usepackage{hyperref}% add hypertext capabilities
\usepackage{epstopdf}
\usepackage{color}
\usepackage{booktabs}
\begin{document}

\preprint{APS/123-QED}

\title{A heat pump without particle transport or external work on the medium achieved by differential thermostatting of the phase space }% Force line breaks with \\

\author{Puneet Kumar Patra}
\affiliation{%
 Advanced Technology Development Center, Indian Institute of Technology Kharagpur, West Bengal, India 721302 
}%
\author{Baidurya Bhattacharya}%
 \email{baidurya@civil.iitkgp.ernet.in}
\affiliation{
Department of Civil Engineering, Indian Institute of Technology Kharagpur, West Bengal, India 721302 
}

\begin{abstract}
We propose a new mechanism that enables heat flow from a colder region to a hotter region without necessitating either particle transport or external work on the conductor, thereby bypassing the compressor part of a classical heat pump cycle. Our mechanism relies on thermostatting the kinetic and configurational temperatures of the same particle differently. We keep the two ends of a conductor, which in the present study is a single dimensional $\phi^4$ chain, at the same kinetic temperature $T_0$, but at different configurational temperatures - one end hotter and the other end colder than $T_0$. While external energy is needed within the thermostatted regions to achieve this differential thermostatting, no external work is performed on the system itself. We show that the mechanism satisfies the statistical form of the second law of thermodynamics (the fluctuation theorem). The proposed mechanism reveals two interesting findings - (i) contrary to traditional thermodynamics where only the kinetic temperature is thought to govern heat conduction, configurational temperature can also play an important role, and (ii) the relative temperature difference between the kinetic and configurational variables governs the direction of heat flow. The challenge, however, is in developing experimental techniques to thermostat the kinetic and configurational variables of the same particle at different values. 

%\begin{description}
%\item[PACS numbers]
%05.10.-a, 05.45.Pq
%\end{description}
\end{abstract}

% PACS, the Physics and Astronomy
                             % Classification Scheme.
%\keywords{Suggested keywords}%Use showkeys class option if keyword
                              %display desired
\maketitle

%\tableofcontents

\section{Introduction}
Any useful work extracted from an engine in a finite time, as is true of any real process natural or engineered, must involve a net flow of heat taking place away from equilibrium. Of the four laws of macroscopic thermodynamics, three have a direct bearing on thermal conduction away from equilibrium. The zeroth law helps define thermal equilibrium, the first law quantifies the dissipated heat, and the second law sets the direction of heat flow. The second law, however, is statistical in nature \cite{ref2,ref3}, and assumes its classical form in the thermodynamic limit. There is a finite probability of heat to flow from a colder region to a hotter region that decreases exponentially with system size and time duration \cite{ref1}. Consequently, even for small scale systems, observed over a sufficiently long duration, heat would flow in the usual manner. Classically, heat pumps involve particle transport (``working fluid''), and require external work to be performed on the working fluid itself \cite{ref4}. At small scales particle transport may be eliminated \cite{heat_pump2} but so far external work on the medium itself has not been avoided \cite{heat_pump1,heat_pump2, heat_pump3,heat_pump4}. In the present work, we propose a heat pump that eliminates both particle transport and external work on the medium. The pumping action is achieved by selective thermostatting of the configurational and kinetic variables, in a manner that a difference is created only between the configurational temperatures at the two thermostatted ends (keeping the kinetic temperatures equal). The thermostats need energy to maintain the selective temperature differences, however, no work is done on the medium itself. The proposed mechanism, as we will show later, is consistent with the second law of thermodynamics, and due to the elimination of the external work on the medium, can lead to the development of newer energy efficient devices. 

Until now, either of the kinetic ($T_K$) or the configurational temperature ($T_C$), 
\begin{equation}
\begin{array}{cc}
T_{K} = \left\langle \dfrac{p_i^2}{2m_i} \right\rangle_e, &
T_{C} = \dfrac{\left\langle | \nabla_x \Phi \left(x \right)|^2 \right\rangle_e}{\left\langle \nabla_x^2 \Phi \left( x \right) \right\rangle_e}. \\ \
\end{array}
\label{eq:Temperatures}
\end{equation}
has been controlled in simulations through one of the many non-Hamiltonian thermostats \cite{ref16,ref17,ref38,ref37}. In (\ref{eq:Temperatures}), $x_i$, $p_i$ and $m_i$ represent the position, momentum and mass of the $i^{\text{th}}$ particle, $\Phi(x)$ represents the total potential energy of the system and $\langle . \rangle_e$ represents the average computed over all the particles of the system. In traditional molecular dynamics, ``temperature'' is used interchangeably with kinetic temperature. However, recent simulation studies have shown that controlling the configurational temperature has advantages over the kinetic temperature control in certain nonequilibrium cases like shear flow \cite{ref35}. In fact for dense fluids the configurational part of temperature is more important \cite{ref36}. Recent measurement of configurational temperature using experimental setups \cite{ref32, ref33, ref34} suggests that its applicability lies beyond the confines of theoretical statistical mechanics. But experimental techniques have not matured enough to \textit{control} the configurational temperature.

It has recently been shown that a heat flow can be induced using \textit{only} non-Hamiltonian thermostats \cite{ref10}, and that  Hamiltonian thermostats (both kinetic and Landau-Lifshitz isoconfigurational \cite{ref31}) fail to generate a nonequilibrium steady-state \cite{ref31}. It must be noted, however, that the application of these non-Hamiltonian thermostats results in the ``usual'' heat flow, and they cannot simultaneously control both the kinetic and configurational temperatures at different values. Controlling one temperature leads to an automatic adjustment of the other and hence, a temperature difference between the kinetic and configurational variables cannot be established. The contributions of kinetic and configurational temperatures towards nonequilibrium thermal conduction has remained an open problem until now \cite{ref36}. This question can be answered \textit{only} by thermostatting the kinetic and configurational temperatures at different values.

In the present study, we are able to differently thermostat the kinetic and configurational variables due to the thermostat (PB thermostat) recently developed by us \cite{ref5}. The PB thermostat utilizes all degrees of freedom for controlling the temperature of the system. This paper is organized as follows: we first highlight the PB thermostat and its ability to differently thermostat the kinetic and configurational variables. Next, we detail the mechanism for obtaining heat flow from a relatively colder region to a hotter region. Subsequently, we present our results on the one-dimensional $\phi^4$ chain.

\section{The PB Thermostat}
The PB thermostat enforces the simultaneous control of both the kinetic and configurational temperatures, shown in  (\ref{eq:Temperatures}) \cite{ref5}. It is both deterministic and time-reversible. The governing equations of motion are: 
\begin{equation}
\begin{array}{rcl}
\dot{x_i} & = & p_i - \xi \nabla_{x_i} \Phi, \\

\dot{p_i} & = & - \nabla_{x_i} \Phi - \eta p_i, \\

\dot{\eta} & = & \dfrac{1}{M_\eta} \sum \limits_{i=1}^{3N} \left( p_i^2 - T_K \right), \\

\dot{\xi} & = & \dfrac{1}{M_\xi} \sum \limits_{i=1}^{3N} \left( \left( \nabla_{x_i} \Phi \right)^2 - T_C \nabla_{x_i}^2 \Phi  \right). \\
\end{array}
\label{eq:PBT}
\end{equation}
where, $M_i$ is the mass of the $i^{\text{th}}$ reservoir $(i=\xi,\eta)$ and $N$ is the number of particles in the system. We have assumed that the particles have unit mass and $k_B=1$. The Nos\'e-Hoover \cite{ref16} kinetic thermostat and the Braga-Travis \cite{ref38} configurational thermostat can be obtained from equations of motion \ref{eq:PBT} by substituting $\xi=\dot{\xi} =0$ and $\eta = \dot{\eta} = 0$, respectively. Thus, a PB thermostat may be viewed as the coupling of a Nos\'{e}-Hoover and a Braga-Travis thermostat using two independent reservoirs. The equations of motion have been derived by solving the Liouville's continuity equation in the extended (6N + 2 dimensional) phase-space assuming that the dynamics is ergodic, and the extended phase-space follows a canonical distribution.

Augmented with switching functions, these equations (\ref{eq:PBT}) can simulate a thermal conduction process. Rather than two, the equations of motion now need four thermostat variables:

\begin{equation}
\begin{array}{rcl}
\dot{x_i} & = & p_i - S^L_i \xi^L \nabla_{x_i} \Phi - S^R_i \xi^R \nabla_{x_i} \Phi, \\

\dot{p_i} & = & -\nabla_{x_i} \Phi -  S^L_i \eta^L p_i -  S^R_i \eta^R p_i, \\

\dot{\eta}^L & = & \dfrac{1}{M_\eta} \sum \left(  S^L_i p_i^2 - T_K^L \right), \\

\dot{\eta}^R & = & \dfrac{1}{M_\eta} \sum \left(  S^R_i p_i^2 - T_K^R \right), \\

\dot{\xi}^L & = & \dfrac{1}{M_\xi} \sum \left(  S^L_i \left( \nabla_{x_i} \Phi \right)^2 - T_C^L \nabla_{x_i}^2 \Phi  \right). \\

\dot{\xi}^R & = & \dfrac{1}{M_\xi} \sum \left(  S^R_i \left( \nabla_{x_i} \Phi \right)^2 - T_C^R \nabla_{x_i}^2 \Phi  \right). \\
\end{array}
\label{eq:PBT_conduction}
\end{equation}
Here, $S_i^L$ (or $S_i^R$) denotes the left (or right) switching function which takes up a value of 1 when the $i^{\text{th}}$ particle is in the left (or the right) thermostatted region, and is zero otherwise. A traditional thermal conduction could be simulated by keeping $T_C^L = T_K^L = T^L> T_C^R = T_K^R = T^R$. The significance of this thermostat is in its ability to set the targets $T_C^i$ and $T_K^i$ independently and arbitrarily at any pair of equal \textit{or} unequal values unlike in the other thermostats where we do not have explicit control of thermostatting the kinetic and configurational variables differently. This ability of the PB thermostat enables us to study the relative contributions of the kinetic and configurational variables, as well as engender thermal transport \textit{along} the temperature gradient without necessitating external work.

\section{The Mechanism and Simulation Model}
\begin{figure}
\includegraphics[scale=0.325]{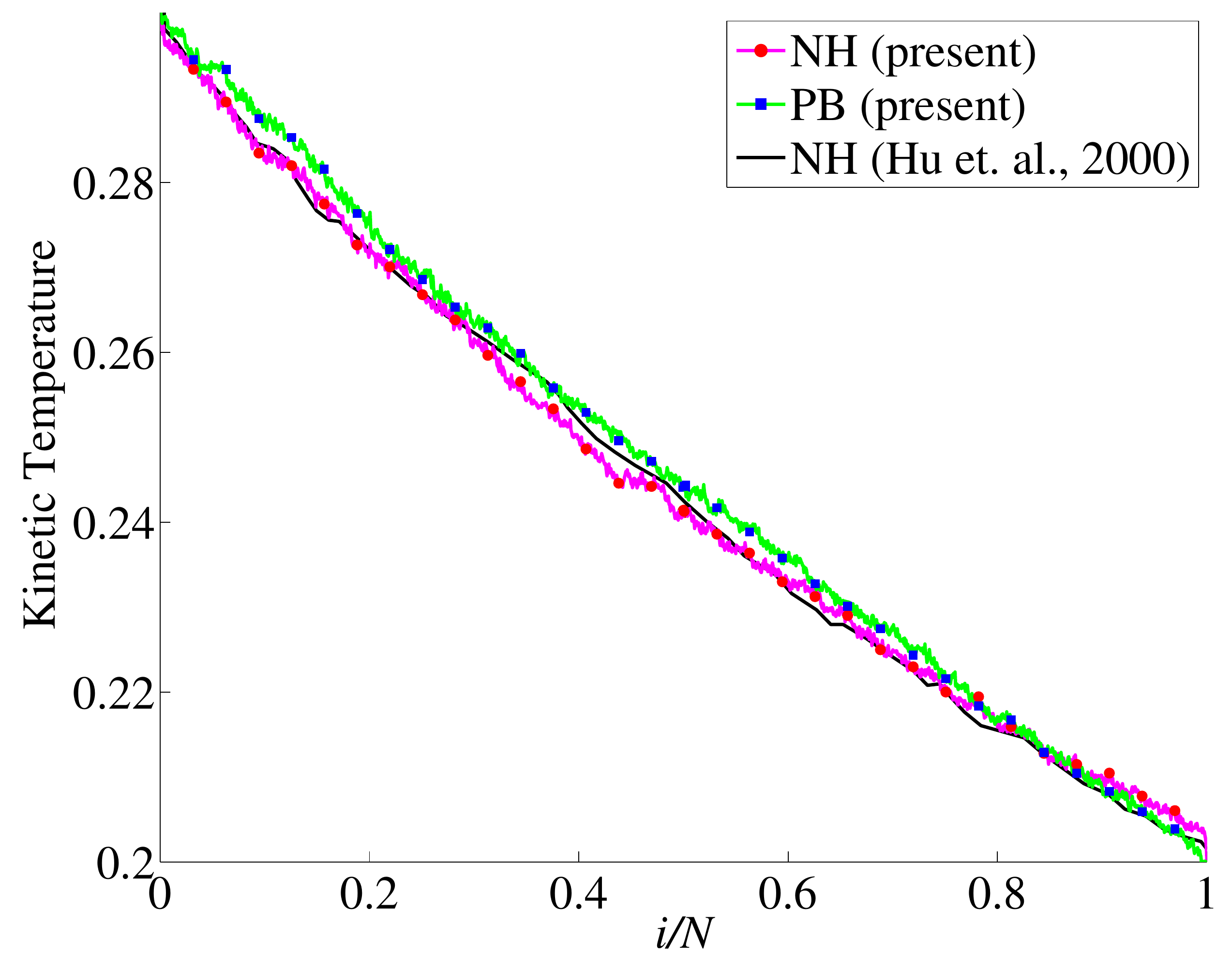}
\caption{\label{fig:bambi_hu} Simulation Results Verification: We compare the kinetic temperature profile due to the NH and PB thermostats with that obtained from Hu \textit{et. al.} (uses NH thermostat)\cite{Hu_2000}. The leftmost particle is thermostatted at a temperature of 0.3, while the rightmost particle is thermostatted at a temperature of 0.2. Overall, there are $N=1600$ particles in the system. The kinetic temperature profiles from our simulation codes reproduce the kinetic temperature profile due to Hu \textit{et. al.} (2000) in a good manner.}
\end{figure}

Let us describe the simulation model adopted in the present study. The system chosen is the prototypical one-dimensional $\phi^4$ thermal conduction model \cite{ref6,ref7,ref8}, which is a nonintegrable system \cite{ref9}, obeys Fourier's law, and has a finite temperature-dependent thermal conductivity of $\kappa = 2.83/T^{1.35}$ \cite{ref10}. In the absence of any thermostatting, the particles of this one-dimension chain are governed by the Hamiltonian
\begin{equation}
\begin{array}{ccc}
H & = & \sum \limits_{i=1}^N \dfrac{p_i^2}{2m_i} + \sum\limits_{i=1}^{N-1}U\left(x_i,x_{i+1} \right) + \sum\limits_{i=1}^N V(x_i)
\end{array}.
\label{eq:phi4_hamiltonian}
\end{equation}
$U(x_i,x_{i+1})$ represents a quadratic nearest neighbour interparticle interaction and is given by $U(x_i,x_{i+1}) = 0.5k_1 \left( | x_{i+1} - x_i | - 1 \right)^2$, while $V(x_i)$ represents the quartic tethering potential and is given by $V(x_i) = 0.25 k_2 \left( x_i - x_{i,0} \right)^4$, with $x_{i,0}$ being the equilibrium position of the $i^{th}$ particle. Here, we choose $k_1=k_2 = 1.0$ and $m_i = 1.0$. To study thermal conduction, the left end of the chain is kept at a higher temperature and the right end is kept at a lower temperature. As stated above, in this ``traditional model'', the kinetic and configurational temperatures are kept such that:  $T_C^L = T_K^L = T^L$ and $T_C^R = T_K^R=T^R$. The traditional model of thermal conduction has been studied by several researchers (with / without explicit configurational temperature control) \cite{ref5,ref9,ref10,ref27,ref28}. For example, Hu \textit{et. al.} \cite{ref9} have studied a $\phi^4$ chain comprising of 1600 particles, with the leftmost particle kept at a kinetic temperature of 0.3 and the rightmost particle kept at a kinetic temperature of 0.2. The results are similar when (i) the PB thermostat (with both kinetic and configurational temperature control) and (ii) the NH thermostat (obtained from (\ref{eq:PBT_conduction}) by substituting $\xi^i = \dot{\xi}^i=0$) are employed to study the thermal conduction. The results are shown in figure \ref{fig:bambi_hu}. The equations of motion are solved for 200,000,000 time steps, with each time step being equal to 0.01. The kinetic temperature profiles due to the NH and PB thermostats from our code reproduce the kinetic temperature profile of Hu et. al. 
Next, we compare the thermal conductivity obtained from these cases. The  theoretical thermal conductivity is \cite{Aoki_02} $\kappa = 2.83/T^{1.35} = 2.83/0.25^{1.35}=18.39$. Thermal conductivity due to our own simulations are: $\kappa_{NH} = 16.87$ and $\kappa_{PB}=18.69$, and due to Hu et. al. is $\kappa = 15.50$ (approx). Since in the traditional model of thermal conduction we cannot control the kinetic and configurational temperatures differently, the traditional model cannot be used to separately identify the importance of kinetic and configurational variables in thermal conduction.

\begin{figure*}
\includegraphics[scale=0.85]{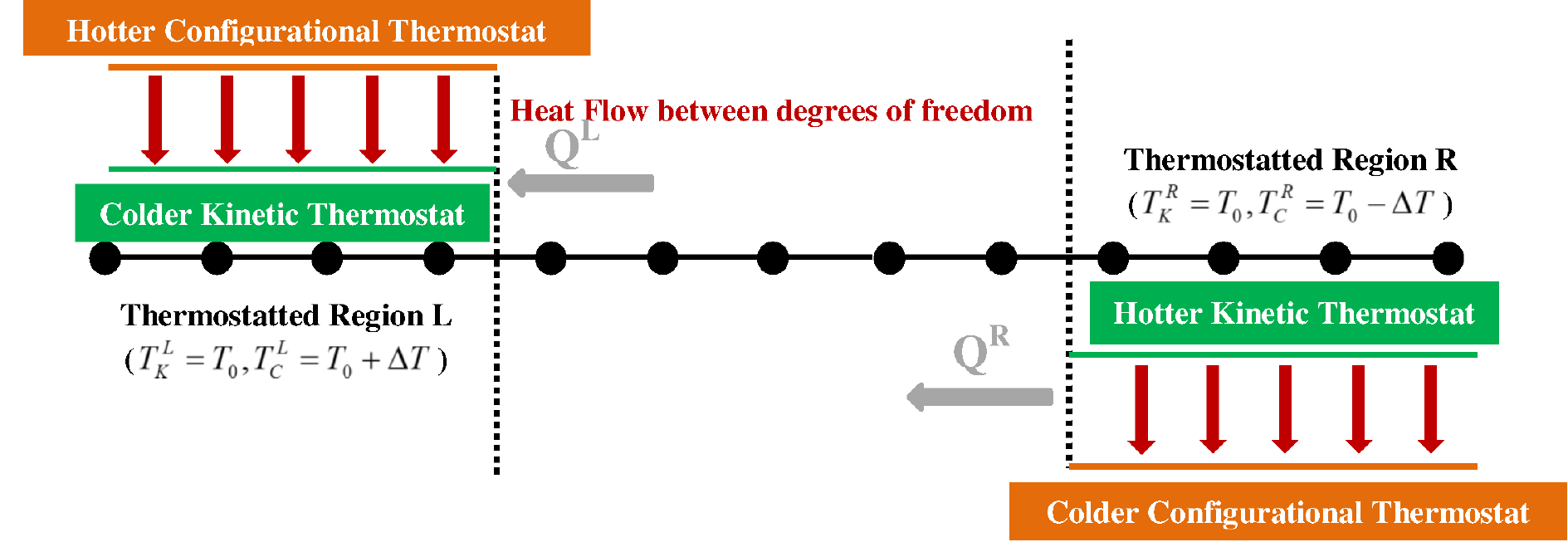}
\caption{\label{fig:FIGURE1} The proposed simulation scheme - kinetic and configurational variables at each thermostatted end are kept at different temperatures that is, the regions $L$ and $R$ are under the influence two PB thermostats. The kinetic temperatures are kept same at both the thermostatted regions ($T_K^L=T_K^R=T_0$). The left thermostatted region has a higher configurational temperature $T_C^L$ than $T_0$ while the right thermostatted region has a configurational temperature $T_C^R$ lower than $T_0$. In steady state, $Q^R$ amount of heat flows from the reservoir $R$ to the system, which is extracted by the reservoir $L$. In steady state, $Q^R \approx Q^L$. }
\end{figure*}

In the present study, rather than keeping $T_C^L = T_K^L$ and $T_C^R = T_K^R$ in equation (\ref{eq:PBT_conduction}), we keep the kinetic and configurational temperatures different at each thermostatted end i.e. $T_C^L \neq T_K^L$ and $T_C^R \neq T_K^R$. For sake of simplicity, we refer to this as the differential thermal conduction model. No previous study has attempted to understand the differential thermal conduction model on any system. The differential thermostatting scheme adopted in this study is shown in Figure (\ref{fig:FIGURE1}). The two ends of the chain are under the influence of two PB thermostats. The intermediate region (comprising of $N_I$ particles) is not under any temperature control and the particles evolve through the usual Hamilton's equations:  $\dot{x_i} = p_i, \dot{p_i}=F_i$. We thermostat the left region (comprising of $N_L$ particles) at a configurational temperature $T_C^L$, and at a kinetic temperature $T_K^L$. Likewise the right region (having $N_R$ particles), is thermostatted at $T_C^R$ and $T_K^R$, respectively. A temperature difference is then created across the chain in a manner that $T_C^L>T_K^L=T_0=T_K^R>T_C^R$, i.e. a temperature difference is imposed only amongst the configurational variables. In the present study, $2T_0=T_C^L+T_C^R$.

We solve the $2N+4$ equations of motion (shown in (\ref{eq:PBT_conduction})) using the classical fourth order Runge-Kutta method for different values of $N$. The equations of motion (with $M_{\xi}=M_{\eta} = 1/1000$) are solved for 500,000,000 time steps, each of size 0.01. The system is first equilibrated at a temperature of 1 for 10,000,000 timesteps. Averages are calculated using the last 250,000,000 time steps. Simulations have been performed for $N = 50$ to $1000$ particles, and $\Delta T = 0.5 (T_C^L - T_C^R) = 0.05$ to $0.30$. For cases with $N>500$, we limit ourselves to $\Delta T = 0.10, 0.20, 0.30$ because of computational requirements. The kinetic temperatures at both the thermostatted ends are kept at 1 i.e. $T_K^L=T_K^R=1$. The configurational temperatures at the thermostatted ends follow the relation: $T_C^L=1+\Delta T$ and $T_C^R=1-\Delta T$. 20\% of the total particles at each ends are under the influence of the thermostats.

To show that the heat flows from right (relatively colder region) to  left (relatively hotter region) we will utilize the facts that (i) the heat flux $J<0$ for heat flow from the hotter region to the colder region, and $J>0$ otherwise, and (ii) a net heat is supplied by the right thermostatted region, which is then extracted by the left thermostatted region. 

Let us now look how we can calculate these thermodynamic variables. The average energy current from the $(i-1)^{th}$ particle to the $i^{th}$ particle for the ones present in the intermediate region is \cite{ref27,ref28}:
\begin{equation}
\langle j_{i,i-1} \rangle_t = \left\langle \dfrac{1}{2} \left( v_i + v_{i-1} \right) \dfrac{\partial U(x_{i-1},x_i)}{\partial x_i} \right\rangle_t.
\end{equation}
$\langle . \rangle_t$ indicates long time averaged value. In steady state, the energy current between any two neighbouring particles must be same, and  the heat flux may be written as:
\begin{equation}
\begin{array}{ccc}
J & = & \dfrac{1}{N_I} \left\langle \sum\limits_{i=N_L}^{N_L+N_I}j_{i,i+1} \right\rangle_t.
\end{array}
\label{eq:heat_current}
\end{equation}

The cumulative heat exchange with the hot and cold thermostatted reservoirs, denoted by $Q^L$ and $Q^R$ respectively, can be calculated by integrating the rate \cite{ref29},
\begin{equation}
\dot{Q}^{i} = \dot{Q}_K^{i} + \dot{Q}_C^{i} = -\left[ \sum_{j=1}^{N_i} \eta^{i}p_j^2 + \sum\limits_{j=1}^{N_{i}} \xi^{i} \left( \dfrac{\partial\Phi}{\partial x_j} \right)^2 \right],
\label{eq:net_heat_flow}
\end{equation} 
with $i=L,R$ depending on the region and $\Phi=U+V$. Associated with $J < 0$ is $Q^L > 0$ and $Q^R<0$, while for $J > 0$, we have $Q^L < 0$ and $Q^R > 0$. The latter implies that the heat is supplied from the right thermostatted region, which is then extracted by the left thermostatted region.

\section{Results and Discussions}
\subsection{Temperature Profiles} \label{sec:b}

\begin{figure}
\includegraphics[scale=0.325]{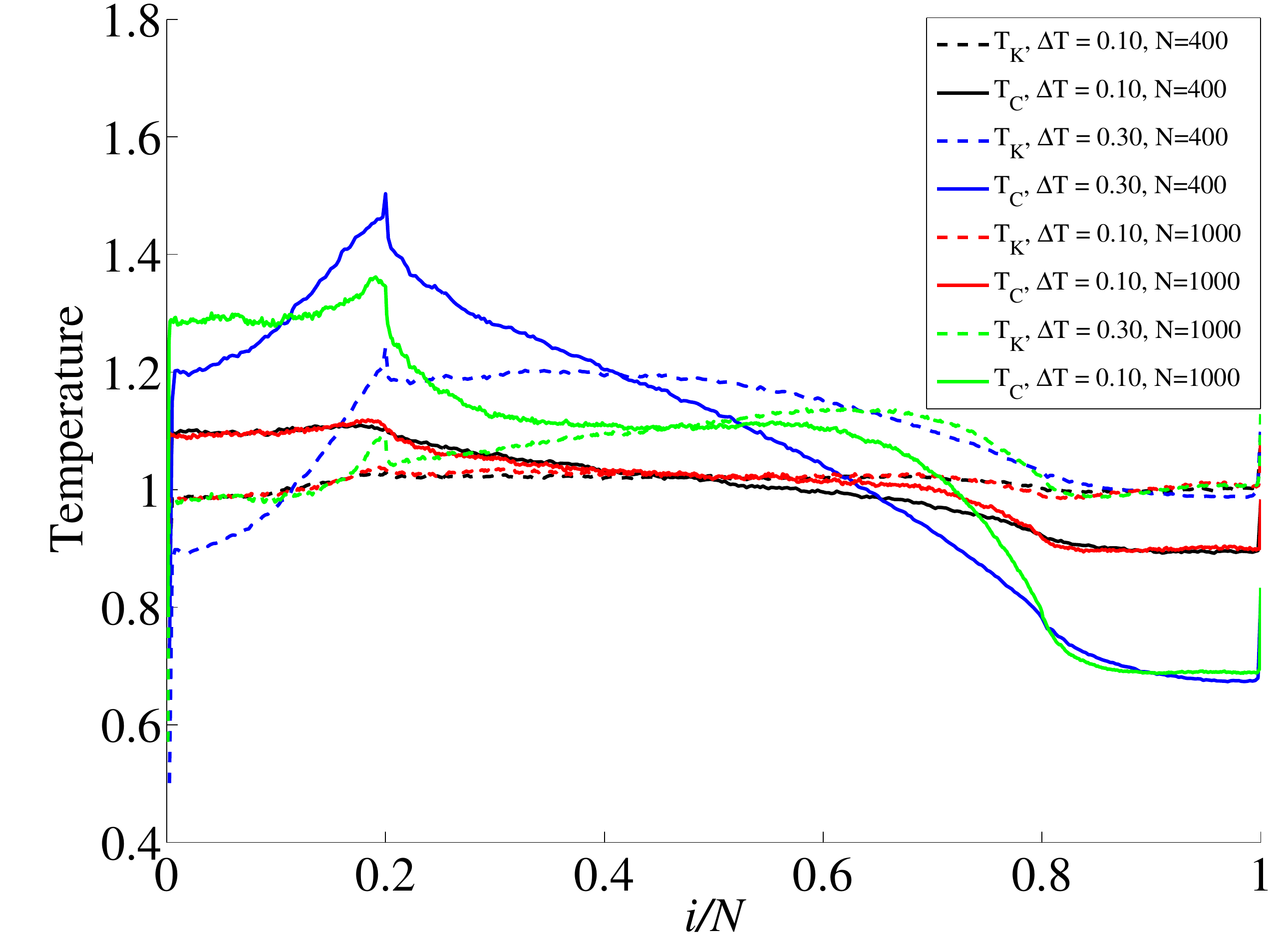}
\caption{\label{fig:temperature_profile_a} The kinetic (dotted lines) and configurational (solid lines) temperature profiles due to $N=400$ and 1000. 20\% of the particles thermostatted at each end. The results are for $\Delta T = \left( T_C^L - T_C^R \right)/2 = 0.20$ and $T_0= 1$. The averages are computed using the last 250 million timesteps. The configurational temperature drops in the intermediate region, as expected. In contrast, the kinetic temperature, despite being maintained at 1 at both ends, bulges up in the intermediate region. Both kinetic and configurational temperature profiles are \textit{asymmetric}. The asymmetry increases with increasing system size. It is evident that the left end, on an average, is hotter than the right end.}
\end{figure}

Let us first establish that (i) the differential thermostatting model puts the system out of local thermodynamic equilibrium, and (ii) the left end is at a higher temperature than the right end. To show (i) we plot the particle wise kinetic and configurational temperature profiles in figure \ref{fig:temperature_profile_a} for $\Delta T = 0.10$ and 0.30 with $N = 400$ and 1000. The temperature profile has a high dependence on $N$. Regardless, a common feature can be observed for all cases -- in the unthermostatted middle region, $T_K$ and $T_C$ are not equal \textit{locally} no matter how small $\Delta T$ and value of $N$ are. Thus, our mechanism puts the system out of local thermal equilibrium. The violation of local thermodynamic equilibrium becomes more pronounced when $\Delta T$ increases. 

\begin{figure*}
\includegraphics[scale=0.325]{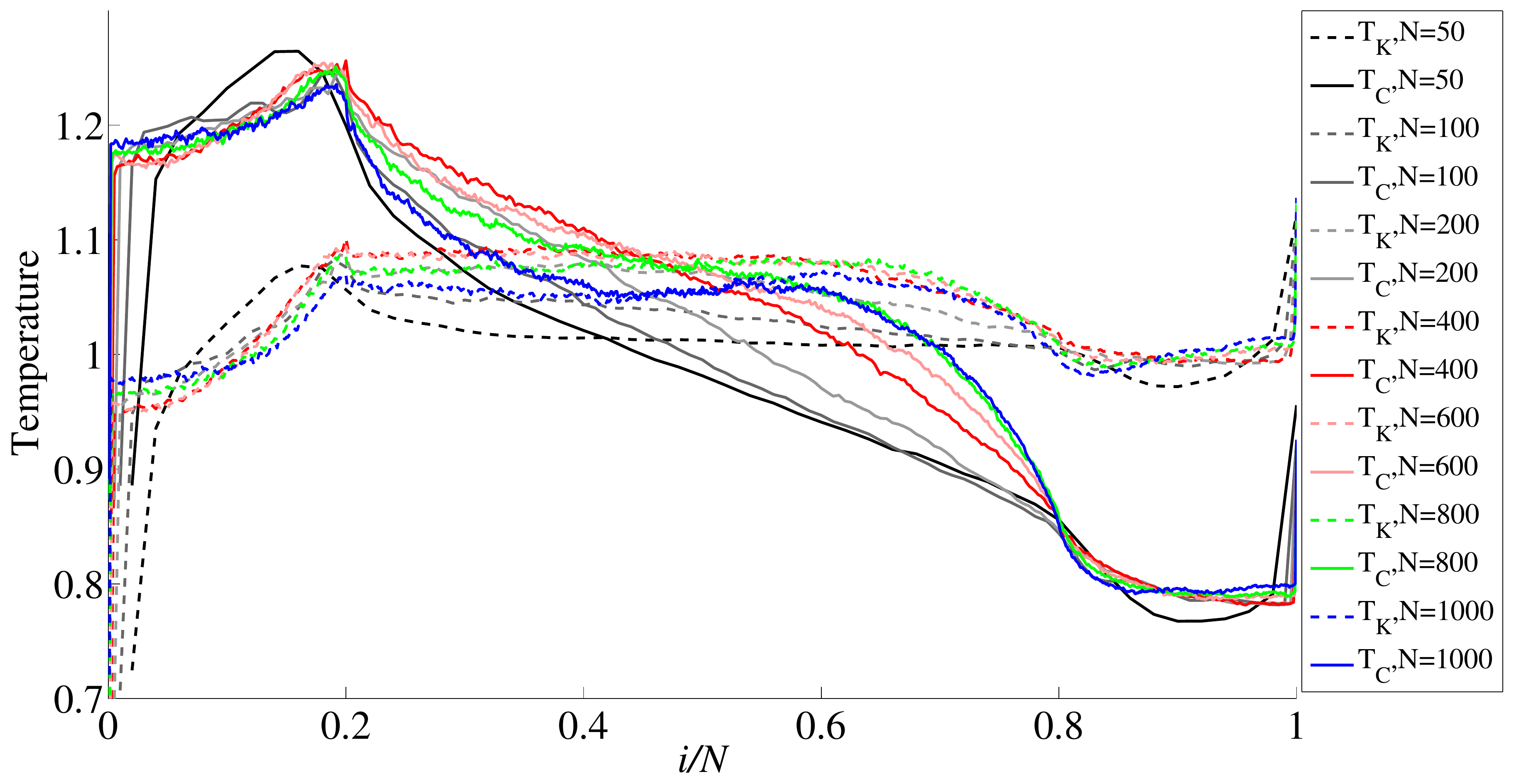}
\caption{\label{fig:temperature_profile_b} Dependence of temperature profiles on system size $N$. These results are for $\Delta T = 0.20$. The asymmetry we saw in the previous figure can be seen clearly in this figure. It is interesting to know that the difference between the kinetic and configurational temperatures \textit{locally} decreases with increasing $N$. }
\end{figure*}

An in-depth look at the temperature profile dependence on $N$ can be seen from figure \ref{fig:temperature_profile_b}. Like before, we observe a significant difference between $T_K$ and $T_C$ locally. Moreover, this difference decreases with increasing $N$.

The average temperature (i.e. $0.5(T_C+T_K)$) indicates that the left end is hotter than the right end, as may be expected. The configurational temperature drops in the intermediate region which also is expected since a gradient has been imposed in $T_C$, and its trend follows the average temperature profile. Interestingly, although $T_K$ is kept same at both ends, it bulges up in the middle, suggesting that the absolute velocities of the particles at this region are higher than at the ends. More interestingly, the profile of neither temperature is symmetric across the conductor. In the presence of the asymmetry and increased velocity of the middle particles, the heat flux turns out to be positive -- that is, heat flows from the lower configurational temperature region to the higher configurational temperature region, which is the central finding of this paper.

\subsection{Equivalence of entropy production and phase-space compression} \label{sec:c}
One of the important properties of the traditional model of thermal conduction is the equivalence of the thermodynamic dissipation as described by (i) heat transfer entropy production ($\dot{S}$) and (ii) the phase-space volume loss ($\Lambda$) \cite{patra2015equivalence}:
\begin{equation}
\langle \dot{S} \rangle_t \equiv \langle \dfrac{\dot{Q}^L}{T^L}+\dfrac{\dot{Q}^R}{T^R} \rangle_t = \langle \dfrac{\partial \dot{x}}{\partial x} + \dfrac{\partial \dot{p}}{\partial p} \rangle_t \equiv \langle \Lambda \rangle_t,
\label{eq:entropy_production_1}
\end{equation}
An important finding of the present work is the validity of (\ref{eq:entropy_production_1}) but now with \textit{individual terms arising due to each thermostat}:
\begin{equation}
\langle \dot{S} \rangle_t \equiv \left\langle{\dfrac{\dot{Q}_K^L}{T_K^L}+\dfrac{\dot{Q}_K^R}{T_K^R}+\dfrac{\dot{Q}_C^L}{T_C^L}+\dfrac{\dot{Q}_C^R}{T_C^R}}\right\rangle_t = \langle \Lambda \rangle_t,
\label{eq:entropy_production}
\end{equation}
where the heat flows $\dot{Q}_i^j$ are according to (\ref{eq:net_heat_flow}) and $\Lambda$ is given by:
\begin{equation}
\begin{array}{rcl}
\Lambda & \equiv & \Lambda_K^L + \Lambda_K^R + \Lambda_C^L + \Lambda_C^R \\
 & = & -N_L\eta^L - N_R \eta^R - \sum\limits_{j=1}^{N_L}\xi^L \dfrac{\partial^2 \phi}{\partial x_j^2} - \sum\limits_{j=1}^{N_R}\xi^R \dfrac{\partial^2 \phi}{\partial x_j^2}.
\end{array}
\label{eq:phase_space_compression}
\end{equation}
In a stricter sense, the equality holds true for every individual terms as well: $\langle \dot{Q}_i^j/T_i^j \rangle_t = \langle \Lambda_i^j \rangle_t$. Numerically, the difference between the two dissipations is negligible, of the order of $10^{-6}$ or smaller (see table \ref{tab:table1}). 

\begin{table*}[]
\centering
\caption{Difference between heat transfer entropy production $\langle \dot{Q_i^j} \rangle_t$ and phase-space volume loss $\Lambda_i^j$ due to the individual four thermostats for $\Delta T = 0.20$. The results suggests that the difference is negligible, and the equality \ref{eq:entropy_production} holds true. Similar results were obtained for all other cases as well.}
\label{tab:table1}
\begin{tabular*}{0.95\textwidth}{@{\extracolsep{\fill}}ccccc@{}}
$N$  & $\langle \dot{Q}_K^L/T_K^L - \Lambda_K^L \rangle_t$        & $\langle \dot{Q}_C^L/T_C^L - \Lambda_C^L \rangle_t$        & $\langle \dot{Q}_K^R/T_K^R - \Lambda_K^R \rangle_t$        & $\langle \dot{Q}_C^R/T_C^R - \Lambda_C^R \rangle_t$ \\ \hline \hline
50	&	-1.00$\times 10^{-7}$	&	-3.40$\times 10^{-7}$	&	7.91$\times 10^{-7}$	&	-5.38$\times 10^{-8}$	\\
100	&	2.27$\times 10^{-8}$	&	2.13$\times 10^{-7}$	&	-5.52$\times 10^{-7}$	&	-2.56$\times 10^{-7}$	\\
200	&	8.33$\times 10^{-7}$	&	2.09$\times 10^{-6}$	&	4.72$\times 10^{-7}$	&	-3.76$\times 10^{-7}$	\\
400	&	-2.02$\times 10^{-8}$	&	3.35$\times 10^{-9}$	&	-8.70$\times 10^{-8}$	&	-6.19$\times 10^{-7}$	\\
600	&	-3.25$\times 10^{-8}$	&	2.46$\times 10^{-8}$	&	6.15$\times 10^{-7}$	&	4.38$\times 10^{-7}$	\\
800	&	5.26$\times 10^{-7}$	&	-5.67$\times 10^{-9}$	&	-1.49$\times 10^{-7}$	&	1.15$\times 10^{-7}$	\\
1000	&	-3.25$\times 10^{-6}$	&	3.81$\times 10^{-7}$	&	9.65$\times 10^{-8}$	&	-5.03$\times 10^{-7}$	\\
 \hline \hline
\end{tabular*}
\end{table*}

The equivalence of equations (\ref{eq:entropy_production}) and (\ref{eq:phase_space_compression}) reaffirms the fact that the thermostats do not perform any work on the system, and are involved only in supplying / withdrawing heat from it, unlike some thermostats \cite{bright_05}. It is remarkable that the equality holds true \textit{despite the violation of local thermodynamic equilibrium} at both the thermostatted and unthermostatted regions. 

We next turn our attention to showing that the heat in the differential thermostatting scheme flows from right to left, i.e. from the relatively colder to the hotter region.

\subsection{Heat Flux and Heat flows} \label{sec:d}

\begin{figure}
\includegraphics[scale=0.325]{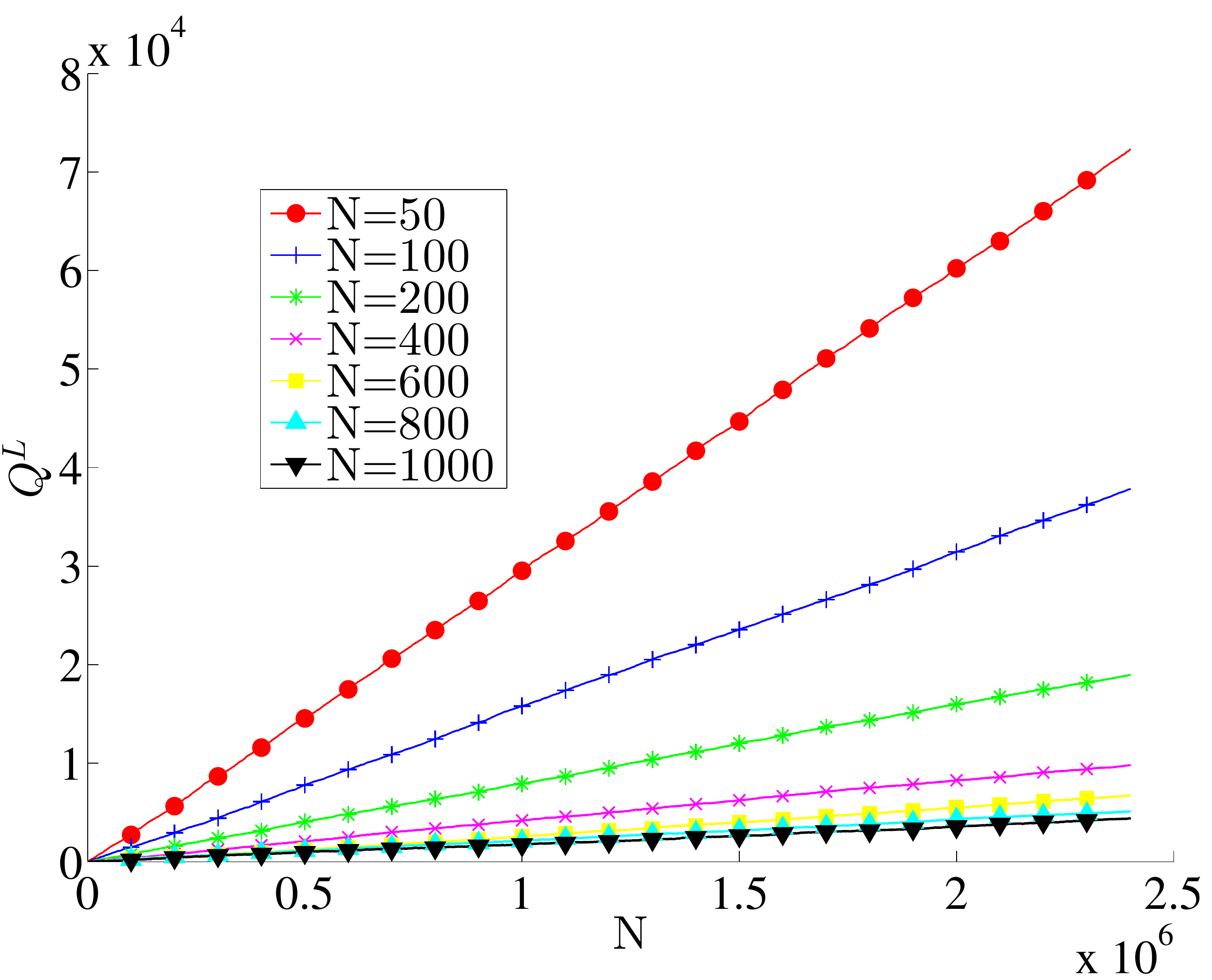}
\caption{\label{fig:heat_flow_both_delT_0.2} Cumulative heat flow from the left thermostatted region for different $N$ under traditional thermostatting with $T_K^L=T_C^L=1.20$ and $T_K^R=T_C^R=0.80$. The results are for last 250 million time steps. The linear nature of the graph indicates that a steady state has been reached. It is interesting to note that as the system size increases, the heat flow from the thermostat decreases. This is consistent with the non-diverging characteristic of $J \times N$ for a $\phi^4$ chain. The positivity of $Q^L$ suggests that the \textit{heat is supplied from the ``hotter'' left thermostatted region} to the system. }
\end{figure}

The proposed scheme allows the heat to flow from the colder region to a hotter region, which can be proved numerically by looking at (i) the sign of $Q^L$ and $Q^R$, and (ii) the sign of $J$. Regardless of the model of thermal conduction (either traditional or differential), for a system to be in steady state, the heat supplied by one of the thermostats must get extracted by the other thermostat, implying that $Q^L+Q^R \approx 0$. Thus, $Q^L \approx -Q^R$, and so we study only $Q^L$. Before going through the results of differential thermostatting scheme, let us take a look at the heat flows during traditional thermostatting scheme with $T_K^L=T_C^L=1.20$ and $T_K^R=T_C^R=0.80$. The results are shown in figure \ref{fig:heat_flow_both_delT_0.2}.

The $\phi^4$ chain has a finite thermal conductivity, which suggests that $J \times N$ a finite value \cite{ref31,Aoki_03,Aoki_00}. In figure \ref{fig:heat_flow_both_delT_0.2} we observe that the heat flowing from the hot thermostat progressively decreases with $N$. This is consistent with the finite thermal conductivity in $\phi^4$ chain. If we had observed that $Q^L$ increases with increasing system size, it would have implied that $J$ also increases with $N$, which in turn would make the thermal conductivity divergent. We also bring to attention that \textit{$Q^L$ is positive} which implies that the \textit{hotter left end supplies heat} to the remainder of the system. 

\begin{figure}
\includegraphics[scale=0.325]{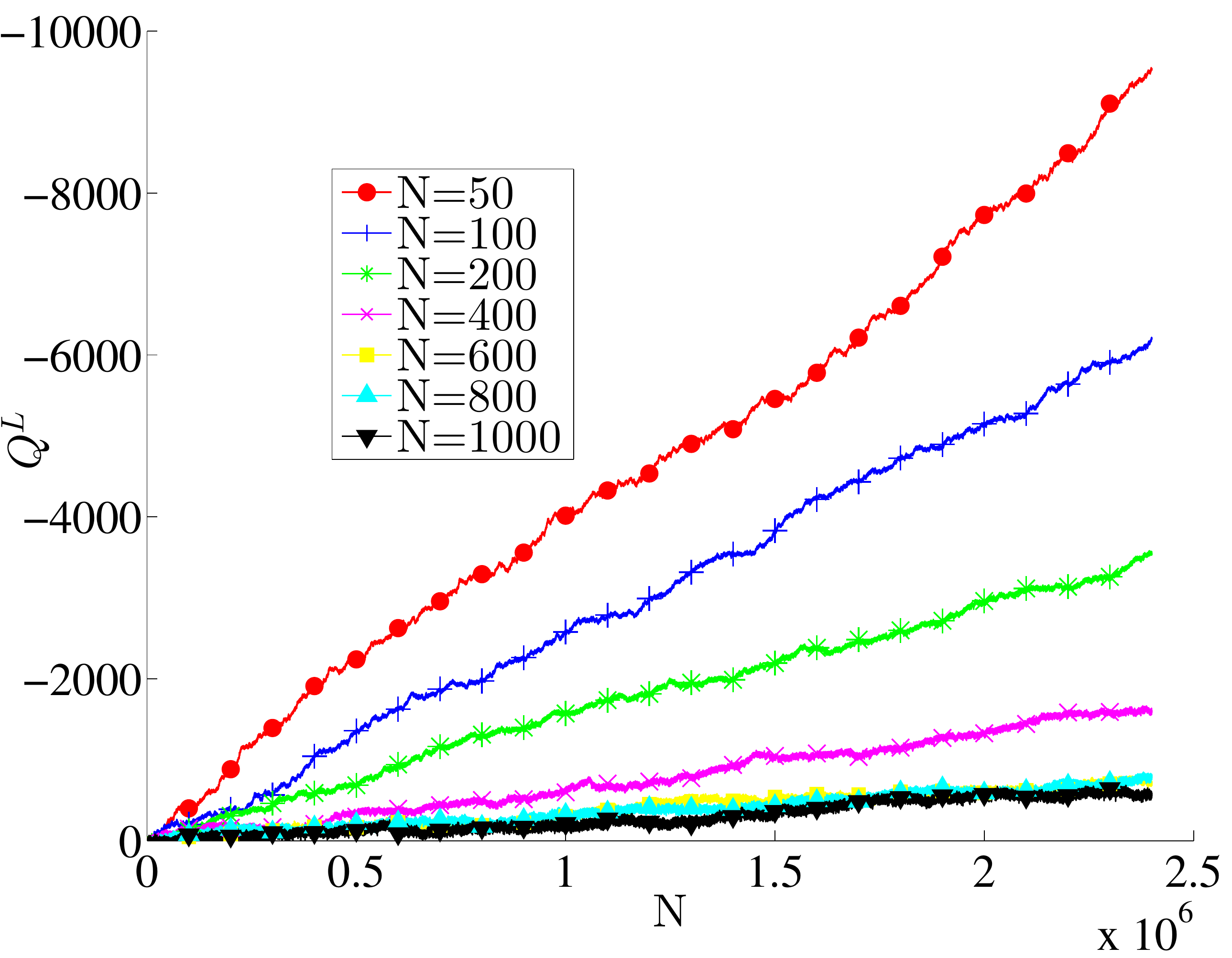}
\caption{\label{fig:heat_flow_no_kinetic_0.2} Cumulative heat flow from the left thermostatted region for different $N$ under differential thermostatting with $T_K^L=T_K^R=1.0$, $T_C^L=1.20$ and $T_C^R=0.80$. The results are for last 250 million time steps. The heat-flows are \textit{almost} linear in nature suggesting that the system is in steady state. Like before, we observe that the heat flow from the thermostat decreases with increasing $N$. Notice that the sign of $Q^L$ is negative, which suggests that the relatively ``hotter'' left thermostatted region \textit{``siphons'' heat from the system}. }
\end{figure}

Now, let us look at the differential thermostatting model with $T_K^L=T_K^R=1.0$, $T_C^L=1.20$ and $T_C^R=0.80$. We remind the readers that \textit{in an averaged sense} the left end of the chain is \textit{hotter} than the right end. The central finding of this paper is $Q^L<0$, as shown in figure \ref{fig:heat_flow_no_kinetic_0.2}, contrary to the normal expectation of $Q^L$ to be positive. The cumulative heat-flows are \textit{almost} linear in nature with no detectable periodicity (the noise appears to be random), suggesting that the system is in steady state. Like before, we observe that the heat flow from the thermostat decreases with increasing $N$. The implication of $Q^L$ being negative is that the relatively ``hotter'' left thermostatted region \textit{withdraws heat from the system} in the same manner as a siphon mechanism. Interestingly this behavior is seen irrespective of the system size and $\Delta T = 0.5 \times (T_C^L - T_C^R)$. This persistent flow of heat from the ``colder" to the ``hotter" region indicates that the proposed mechanism can serve as a heat pump. The significance of our work lies in the fact our heat pump does not require any particle transport and external work on the medium. 

The numerical results show that, for the region $R$, the kinetic reservoir (which is hotter) supplies heat: $Q_K^R > 0$, and the configurational reservoir (which is colder) extracts heat: $Q_C^R < 0$. The cumulative heat flows, however, are different ($|Q_K^R| \neq |Q_C^R|$), and the supply is more than the extracted amount ($|Q_K^R| > |Q_C^R|$). Hence, a net heat flows \textit{into} the system from the right. We attribute this to the slow relaxation of the configurational variables in comparison to the kinetic ones, and as a result, a part of the heat supplied by the kinetic reservoir gets transmitted to the chain before it can get extracted by the configurational reservoir. For the region $L$, the kinetic reservoir is colder, and extracts heat from the system: $Q_K^L < 0$, while the configurational reservoir is hotter, and supplies heat to the system: $Q_C^L > 0$. The extraction is more than supply for $L$, and hence, net heat flows out of the system from $L$. At steady state, the net heat from $L$ and $R$ are equal and opposite in magnitude ($Q^L \approx -Q^R$). 

The heat available for flowing into the system from the thermostat in differential thermostatting scheme is almost an order of magnitude \textit{smaller} than the one available in the traditional thermostatting scheme. The reason may be attributed to the heat flow between the hotter and colder variables \textit{within} the thermostatted regions in the differential thermostat scheme. As a consequence, only a part of heat is available for flowing into the system. In the traditional thermostatting scheme, because of absence of such heat flow between the kinetic and configurational variables at the thermostatted regions, a considerably larger amount of heat is available for flowing into the system. 

\begin{figure}
\includegraphics[scale=0.325]{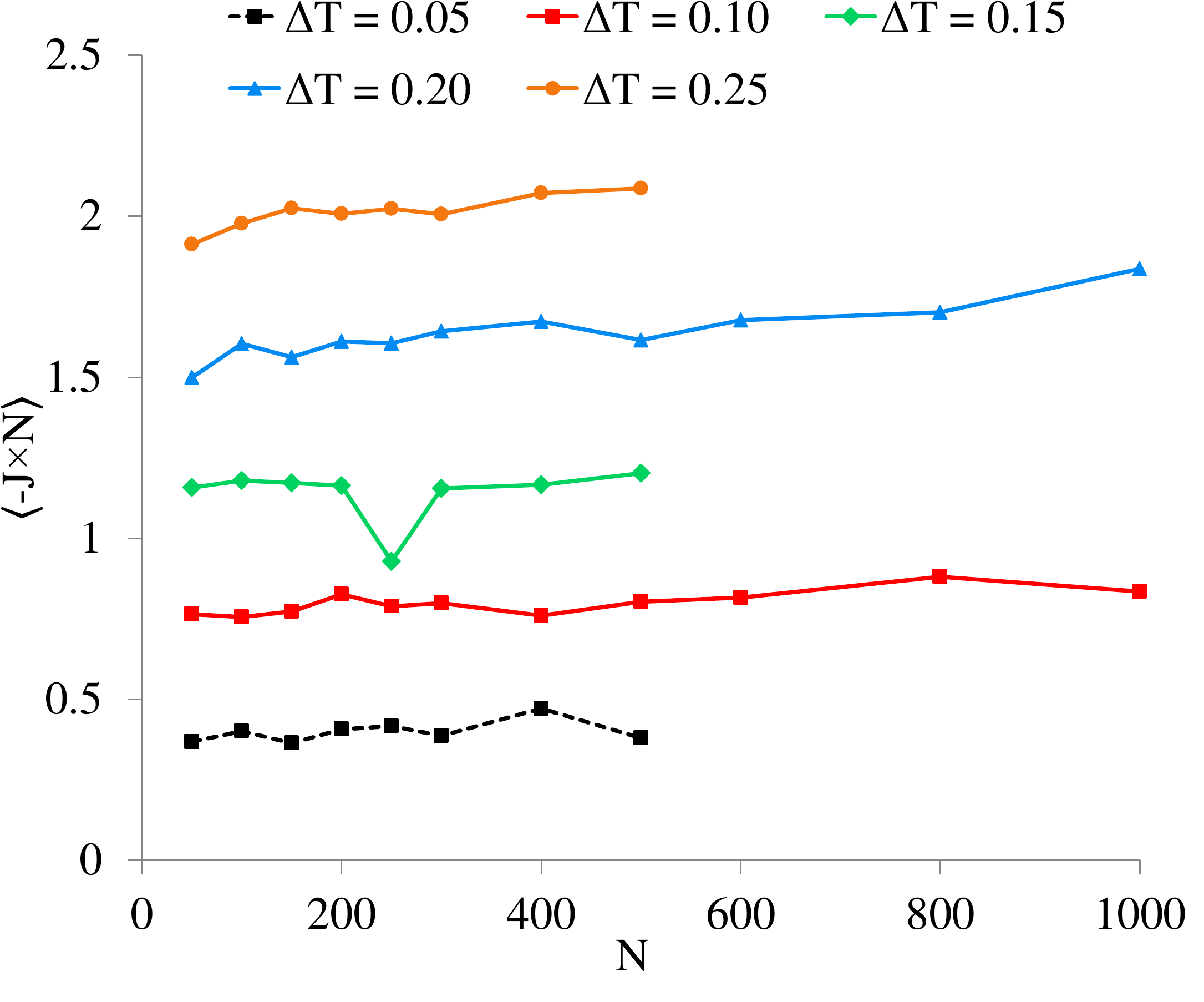}
\caption{\label{fig:heat_flux_figure_both} $-J\times N$ for traditional thermostatting scheme with different values of $\Delta T$ and $N$. Notice that $J < 0$, indicating a heat flow from the hotter left to the colder right region. $|J\times N|$ remains nearly constant with increasing $N$, as expected.}
\end{figure}

Now let us compare the heat flux obtained from the traditional model and the differential model. The heat flux for different values of $\Delta T$ and $N$ with traditional thermostatting are shown in figure \ref{fig:heat_flux_figure_both}. As expected, $J$ is negative suggesting that heat flows from the left hotter region to the right colder region. The heat flux remains fairly constant with increasing $N$. This is consistent with our previous finding that $Q^L$ decreases with increasing $N$.

\begin{figure}
\includegraphics[scale=0.325]{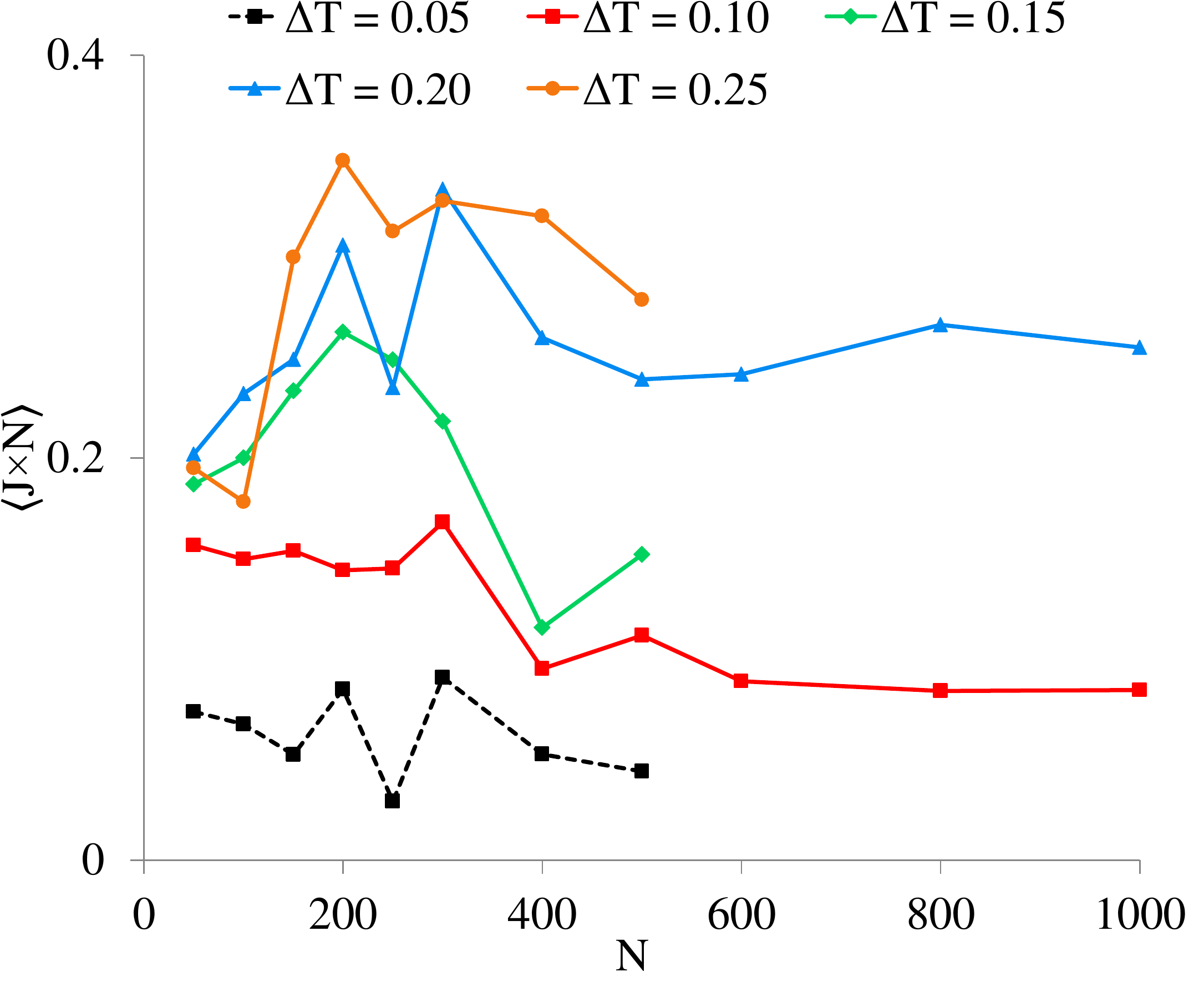}
\caption{\label{fig:heat_flux_figure} $J\times N$ for differential thermostatting scheme with different $\Delta T$ and $N$. Notice that $J$ is positive, suggesting a consistent heat flow from the relatively colder right region to the relatively hotter left region. }
\end{figure}

The heat flux arising due to differential thermostatting are shown in figure \ref{fig:heat_flux_figure}. Notice that $J$ is positive, unlike in the traditional thermostatting scheme. The results indicate a persistent heat flow from the relatively colder right region to the relatively hotter left region. For the majority of the cases the absolute heat fluxes due to the differential thermostatting is an order smaller than the traditional thermostatting. This is because the heat available the thermostatted region (see figures \ref{fig:heat_flow_both_delT_0.2} and \ref{fig:heat_flow_no_kinetic_0.2}) is significantly smaller. 

\subsection{Satisfies Fluctuation theorem for heat flow} \label{sec:e}
 
The results are consistent with the second law of thermodynamics. We use the fluctuation theorem (FT) for heat flow \cite{ref1,ref29,ref30} to demonstrate the second law of thermodynamics in this case:
\begin{equation}
\dfrac{P \left( \langle{\Omega}\rangle_t = A \right)}{P \left( \langle{\Omega}\rangle_t = -A \right)} = \exp \left( At \right),
\label{eq:FTact} 
\end{equation}
where $\langle{\Omega}_t\rangle$ is the time averaged dissipation function defined through:
\begin{equation}
\langle{\Omega}\rangle_t t = \int\limits_0^t\Omega(s)ds = \log \left( \dfrac{f(\Gamma(0),0)}{f(\Gamma(t),0)}\right) - \int\limits_0^t \Lambda(s)ds.
\label{eq:dissi_func} 
\end{equation}
In (\ref{eq:dissi_func}), $f(\Gamma(0),0)$ and $f(\Gamma(t),0)$ denote the density functions of two trajectories that begin at the microstates $\Gamma(0)$ and $\Gamma(t)$, respectively.  Assuming ergodic consistency i.e. a trajectory and its time-reversed conjugate trajectory are associated with nonzero probability, it can be shown that the time averaged dissipation function ($\langle{\Omega}\rangle_t$) for our mechanism becomes
\begin{equation}
\begin{array}{cclc}
\langle{\Omega} \rangle_t & = & \dfrac{1}{t}\dfrac{\Delta T}{T_0} \int\limits_0^t \left( \xi_C^R \sum\limits_{N_R}\dfrac{\partial^2 \phi}{\partial x_i^2} - \xi_C^L \sum\limits_{N_L}\dfrac{\partial^2 \phi}{\partial x_i^2}\right)dt,\\

& = & \dfrac{\Delta T}{T_0} \left( \langle{\alpha}_C^R\rangle_t - \langle{\alpha}_C^L\rangle_t \right).
\end{array}
\label{eq:dissipation_func}
\end{equation}
In (\ref{eq:dissipation_func}), $\langle{\alpha}_C^i\rangle_t$ are time averaged values of the integrals. In this particular set up the kinetic phase-space compression factors play no role in the dissipation function (since $\Delta T$ for kinetic variables is zero), and hence (\ref{eq:dissipation_func}) is devoid of these terms. However, it must be pointed out that the kinetic phase-space compression factors are nonzero, and they play an important role in the heat flow process, as has been highlighted in section \ref{sec:c}. Recasting (\ref{eq:FTact}) in terms of $\langle{\alpha}\rangle$, the fluctuation theorem becomes:
\begin{equation}
\dfrac{P \left( \langle{\alpha}_C^R \rangle_t - \langle {\alpha}_C^L \rangle_t = A \right)}{P \left( \langle{\alpha}_C^R \rangle_t - \langle{\alpha}_C^L \rangle_t = -A \right)} = \exp \left( At \Delta T / T_0\right),
\label{eq:FT} 
\end{equation}

\begin{figure}
\includegraphics[scale=0.325]{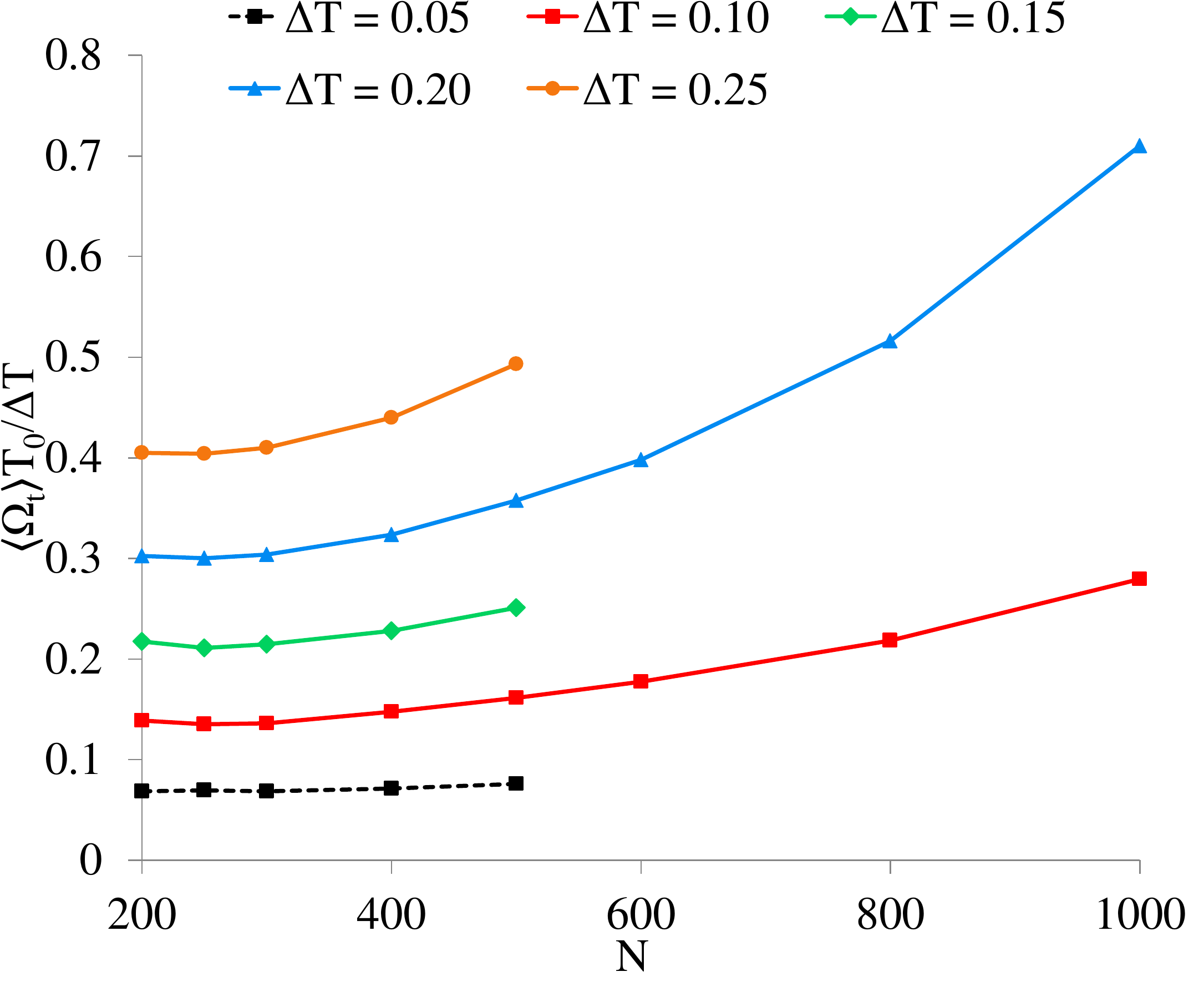}
\caption{\label{fig:phase_space_compression_figure} Variation of $\langle \alpha_C^R \rangle - \langle \alpha_C^L \rangle ( \equiv\langle \Omega_t \rangle T_0 / \Delta T)$ with $\Delta T$ and $N$. It is evident that (i) $\langle \alpha_C^R \rangle - \langle \alpha_C^L \rangle > 0$, (ii) increases with $N$ and (iii) increases with $\Delta T$. Thus the properties of fluctuation theorem are satisfied, and hence, the second law.}
\end{figure}

Thus, for the fluctuation theorem to be satisfied over a long time duration the following must hold true: (i) $\langle{\alpha}_C^R \rangle_t - \langle{\alpha}_C^L\rangle_t > 0$, (ii) $\langle{\alpha}_C^R \rangle_t - \langle{\alpha}_C^L\rangle_t$ must increase with $N$, and (iii) $\langle{\alpha}_C^R \rangle_t - \langle{\alpha}_C^L\rangle_t$ must increase with $\Delta T$. The results shown in figure \ref{fig:phase_space_compression_figure} confirms all these points.

In traditional steady-state heat flow, the hotter thermostat supplies heat to the system causing a phase-space volume expansion, and the colder thermostat must withdraw the same amount of heat causing a phase-space volume compression. It is known however that the steady-state system collapses on an average to a dimension lower than the phase-space dimension, causing a divergence of Gibbs' entropy to negative infinity. It can occur only if the phase-space volume compression due to the colder thermostat exceeds the volume expansion due to the hotter thermostat. Thus, in our problem, for the Gibbs' entropy to diverge: $\langle \Lambda \rangle_t =  \langle \Lambda_K^L+\Lambda_C^L+\Lambda_K^R+\Lambda_C^R \rangle_t= - \langle \sum \alpha_i^j\rangle_t < 0$. Table \ref{tab:table2} shows that $\langle \Lambda \rangle_t < 0$ for different $N$ and $\Delta T$. 

\begin{table}[]
\centering
\caption{Divergence of Gibbs' entropy: Each term of the table denotes $\langle \Lambda \rangle_t$. Notice that $\langle \Lambda \rangle_t < 0$ which suggests that the Gibbs' entropy diverges, a criteria must for nonequilibrium states. Similar values occur for other $N$ and $\Delta T$. }
\label{tab:table2}
\begin{tabular*}{0.45\textwidth}{@{\extracolsep{\fill}}ccc@{}}
$N$  & $\Delta T = 0.10$        & $\Delta T = 0.20$   \\ \hline \hline
200 & -0.0140 & -0.0605 \\
400 & -0.0146 & -0.0653 \\
600 & -0.0177 & -0.0796 \\
800 & -0.0218 & -0.1032 \\
1000 &-0.0279 & -0.1420 \\ \hline \hline
\end{tabular*}
\end{table}

\subsection{Switching the role of kinetic and configurational temperatures}

To judge the relative importance of the different temperatures, we interchange the roles of the configurational and kinetic temperatures in figure \ref{fig:FIGURE1} (i.e. a temperature difference is created only in the kinetic variables). We observed a traditional heat flow (not shown) in this case (i.e. the heat flows from the hotter left region to the colder right region). It is interesting to note that the heat flux in this case is almost an order of magnitude higher than the one observed for differential thermostatting scheme, suggesting the dominant role of the kinetic variables in thermal conduction.

%\begin{enumerate}
%\item Write about the ``error'' computation in ``heat flow rate'' 
%\end{enumerate}

\section{Conclusions}
In this work, we introduce the differential thermostatting scheme where the kinetic and configurational variables at a thermostatted region are kept at different temperatures. Two such differentially thermostatted regions at the two ends of a chain allow the heat to flow from the relatively colder region to the relatively hotter region, without requiring any additional work to be performed on the system. Our results suggest that the relative temperature difference between the kinetic and the configurational variables (at each thermostatted end of the conductor) determines the direction of heat flow, exploiting which a heat pump may be developed. Using this approach, we are able to show the potential importance of configurational variables towards thermal conduction. The challenge, however, is in developing experimental techniques to control the kinetic and configurational temperatures of the same particle at different values. 

%\nocite{*}
\bibliography{apssamp}% Produces the bibliography via BibTeX.
\end{document}